\documentstyle[preprint,aps]{revtex}
\tighten

\begin{document}
\preprint{}
\draft
 
\title{The Pauli principle in the soft-photon approach to 
       proton-proton bremsstrahlung}

\author{M.\ K.\ Liou}
\address{Department of Physics and Institute for Nuclear Theory, \\
         Brooklyn College of the City University of New York, \\
         Brooklyn, NY 11210, USA} 
\author{R.\ Timmermans}
\address{Kernfysisch Versneller Instituut,
         University of Groningen, \\ Zernikelaan 25,
         NL-9747 AA Groningen, the Netherlands}
\author{B.\ F.\ Gibson}
\address{Theoretical Division, Los Alamos National Laboratory, \\
         Los Alamos, NM 87545, USA}

\date{\today}

\maketitle

\begin{abstract}
A relativistic and manifestly gauge-invariant soft-photon amplitude,
which is consistent with the soft-photon theorem and satisfies the
Pauli Principle, is derived for the proton-proton bremsstrahlung process.  
This soft-photon amplitude is the first two-u-two-t special amplitude
to satisfy all theoretical constraints.  The conventional Low
amplitude can be obtained as a special case.  It is demonstrated that 
previously proposed amplitudes for this process, both the (u,t)
and (s,t) classes, violate the Pauli principle at some level.  The 
origin of the Pauli principle violation is shown to 
come from two sources: (i) For the (s,t) class, the two-s-two-t amplitude
transforms into the two-s-two-u amplitude under the interchange of two
initial-state (or final-state) protons.  (ii) For the (u,t) class, the
use of an internal emission amplitude determined from the 
gauge-invariance constraint alone, without imposition of the Pauli
principle, causes a problem.  The resulting internal emission amplitude 
can depend upon an electromagnetic factor which is not invariant under 
the interchange of the two protons.  
\end{abstract}

\newpage
\section{Introduction}
It has been known since the early work of Low~\cite{Low58} that the 
soft-photon theorem applies to all nuclear bremsstrahlung processes.
This theorem states that, when the total bremsstrahlung amplitude is
expanded in powers of the photon momentum (energy) $K$, the coefficients
of the two leading terms are independent of off-shell effects.
Therefore, the theorem implies that a soft-photon approximation (an 
on-shell approximation based upon the first two terms) should provide a 
good description of any bremsstrahlung process, including
proton-proton bremsstrahlung ($pp\gamma$).  The open question has been 
how to construct a soft-photon amplitude which satisfies all theoretical
constraints.

During the past three decades, a variety of soft-photon amplitudes have
been proposed to describe the ($pp\gamma$) process.  Although most of
these amplitudes are relativistic, gauge invariant, and consistent with
the soft-photon theorem, they violate the Pauli principle at some level.
The requirement of fully satisfying the Pauli principle was heretofore
neglected.  The purpose of this paper is to provide a
derivation of a soft-photon amplitude that not only is consistent
with the soft-photon theorem, is valid relativistically, is
manifestly gauge invariant, but also satifies the Pauli principle.

Recently, a prescription to generate two classes of soft-photon
amplitudes was discussed:  (1) the two-$u$-two-$t$ special ($TuTts$)
amplitudes from the class expressed in terms of the ($u,t$)
Mandelstam variables and (2) the two-$s$-two-$t$
special ($TsTts$) amplitudes from the 
class expressed in terms of the ($s,t$) Mandelstam
variables~\cite{Lio93}.  In Ref.~\cite{Lio93}, simple cases were used
to demonstrate basic ideas and methods.  The two particles involved
in the scattering were assumed to be spinless and to have
different masses and charges.  The elastic scattering amplitude was 
defined as the sum of a direct amplitude and an exchange amplitude.
Under these assumptions, the derived amplitudes are applicable to
a description of bremsstrahlung processes involving the scattering
of two bosons, but not two fermions.  Because the proton is a spin-1/2
particle and the two-proton amplitude must obey the Pauli Principle,
the $pp$ elastic amplitude must be antisymmetric under interchange
of the protons.  That is, for the $pp$ case the scattering
amplitude should be obtained as the direct amplitude minus (not
plus) the exchange amplitude.  Therefore, the $TuTts$ amplitude
derived in Ref.~\cite{Lio93}
is not a proper representation of the $pp \gamma$
process, even though the argument regarding why the $TuTts$-type 
amplitude should be used to describe the $pp$ bremsstrahlung process 
is correct.  Moreover, there is an additional problem which is related 
to the ambiguity in determining the internal emission amplitude.  
Without imposing the fermion antisymmetry requirement, the gauge
invariant condition alone does not yield a unique expression for
the internal amplitude.  This important point, emphasized here,
was not imposed in Ref.~\cite{Lio93}.
As a result, the internal amplitude obtained 
in Ref.~\cite{Lio93} for the nonidentical
particles considered is not a proper choice
for bremsstrahlung processes involving two identical nucleons.
For the case of $pp\gamma$ the violation of the Pauli principle for
the TuTts amplitude introduced in Ref.~\cite{Lio93} is not serious
since such violation is found only in the term of order $K$.

A more realistic $TuTts$ amplitude for the $pp\gamma$ process was
proposed recently \cite{Lio95}.  That amplitude is relativistic,
gauge invariant, and consistent with the soft-photon theorem.  However, 
it does not obey the Pauli principle at the $K^1$ order in the 
expansion in terms of $K$.  The problem arises from the internal 
amplitude.  It involves an 
electromagnetic factor which is not invariant under the 
interchange of the two initial-state (incoming) or two
final-state (outgoing) protons.  As we demonstrate below, this
factor is but one of two possible choices that can be obtained 
by imposing gauge invariance.  The second choice for the invariant 
factor was missed in Ref.~\cite{Lio95},
because the requirement that the Pauli principle
be satisfied was not imposed in the derivation.

Except for the $TuTts$ amplitude discussed in Ref.~\cite{Lio95}, almost
all $pp\gamma$ soft-photon amplitudes considered in the literature
belong to the ($s,t$) class.  These amplitudes depend upon the
$pp$ elastic amplitude, which is evaluated at the square of
the total center-of-mass energy $s$ and the square of the momentum 
transfer $t$.  In fact, in most cases the average $s$ and the
average $t$ were used.  The amplitudes obtained by Nyman~\cite{Nym68} 
and Fearing~\cite{Fea80} are two well known examples.  
Such amplitudes are classified as Low amplitudes.  Except for
the Low amplitudes, all other amplitudes in the (s,t) class violate
the Pauli principle for the following reason:
If one interchanges the two initial-state (or final-state) protons,
then one converts the ($s,t$) class of amplitudes into the ($s,u$)
class of amplitudes.  Because the $pp\gamma$ process involves a
half-off-shell amplitude (not an elastic ampltude), the ($s,u$)
amplitude obtained by this procedure is completely different from 
the original ($s,t$) amplitude.  Therefore, it is impossible
to regain the original ($s,t$) amplitude with just a sign change
after interchanging the two protons.

This paper is structured as follows.  In Sec.\ II we define the
$pp$ elastic scattering amplitude which will be used as input to
generate the bremsstrahlung amplitudes for the $pp\gamma$
process.  We use the amplitude introduced by Goldberger, Grisaru, 
MacDowell, and Wong (GGMW)~\cite{Gol60}, but without incorporating 
the Fierz transformation.  In Sec.\ III we derive a relativistic
$TuTts$ amplitude by imposing gauge invariance and the Pauli
principle.  In deriving the amplitude, a straightforward and
rigorous approach, slightly different from that employed in 
Ref.~\cite{Lio93}, is utilized.  We verify that the resulting $TuTts$ 
amplitude is consistent with the soft-photon theorem.  Finally,
a variety of other amplitudes, which violate the Pauli principle,
are discussed in Sec.\ IV.

\section{The proton-proton elastic scattering amplitude}
%---------------------------------------------------------
The Feynman amplitude $F$ for $pp$ elastic scattering,
\begin{equation}
   p(q^{\mu}_i) + p(p^{\mu}_i) \longrightarrow p(\bar{q}^{\mu}_f) +  
   p(\bar{p}^{\mu}_f) \ ,
\end{equation}
can be written as~\cite{Gol60}
\begin{eqnarray}
   F &=& F_1 (G_1 - {\stackrel{\textstyle \sim}{G}}_1) 
       + F_2 (G_2 + {\stackrel{\textstyle \sim}{G}}_2) 
       + F_3 (G_3 - {\stackrel{\textstyle \sim}{G}}_3)
       + F_4 (G_4 + {\stackrel{\textstyle \sim}{G}}_4) 
       + F_5 (G_5 - {\stackrel{\textstyle \sim}{G}}_5) \nonumber \\
     &=& \sum^5_{\alpha = 1} F_{\alpha} [G_{\alpha} + (-1)^{\alpha}  
         {\stackrel{\textstyle \sim}{G}}_{\alpha}] \ ,
\end{eqnarray}
where
\begin{eqnarray}
   G_\alpha &=& \bar{u} (\bar{q}_f) \lambda_{\alpha} u (q_i) \bar{u} 
   (\bar{p}_f) \lambda^\alpha u(p_i) \ , \nonumber \\
   {\stackrel{\textstyle \sim}{G_\alpha}} &=& \bar{u} (\bar{p}_f) 
    \lambda_{\alpha} u (q_i) \bar{u} 
   (\bar{q}_f) \lambda^{\alpha} u(p_i) \ ,
\end{eqnarray}
and we define
\begin{eqnarray}
   (\lambda_1,\lambda_2,\lambda_3,\lambda_4,\lambda_5) & \equiv &
   (1,\frac{\sigma_{\mu\nu}}{\sqrt{2}},i\gamma_5\gamma_\mu,
    \gamma_\mu,\gamma_5) \ , \nonumber \\
   (\lambda^1,\lambda^2,\lambda^3,\lambda^4,\lambda^5) & \equiv &
   (1,\frac{\sigma^{\mu\nu}}{\sqrt{2}},i\gamma_5\gamma^\mu,
    \gamma^\mu,\gamma_5) \ . \nonumber
\end{eqnarray}
Note that $\lambda_\alpha$ and $\lambda^\alpha$ are tensors.
For example, $\lambda^2 \lambda_2 = \lambda_2 \lambda^2 =$ 
$\frac{1}{2} \sigma_{\mu \nu} \sigma^{\mu \nu}$, where the
summation over $\mu$ and $\nu$ is implied.
In Eq.\ (2), $F_{\alpha}$ ($\alpha=1,\dots,5$) are invariant
functions of the Mandelstam variables $s$, $t$, and $u$,
\begin{eqnarray}
   s & = & (q_i+p_i)^2 = (\bar{q}_f+\bar{p}_f)^2 \ , \nonumber \\
   t & = & (\bar{p}_f-p_i)^2 = (\bar{q}_f-q_i)^2 \ , \nonumber \\
   u & = & (\bar{p}_f-q_i)^2 = (\bar{q}_f-p_i)^2 \ .
\end{eqnarray}
Because of energy-momentum conservation,
\begin{equation}
   q^{\mu}_i + p^{\mu}_i = \bar{q}^{\mu}_f + \bar{p}^{\mu}_f \ ,
\end{equation}
$s$, $t$, and $u$ satisfy the following relation,
\begin{equation}
   s + t + u = 4 m^2,
\end{equation}
so that only two of them are independent.  (Here $m$ is the proton mass.)
The optimal choice of these two independent variables will
depend on the fundamental diagrams (or the dominant tree
diagrams) of a given process.  In our case, guided by a
meson-exchange theory of the $N\!N$ interaction, we choose $u$ and
$t$ to be the two independent variables, and we write 
$F_{\alpha}=F_{\alpha}(u,t)$. 
In Eq.\ (2), $\sum^{5}_{\alpha=1} F_{\alpha}(u,t)G_{\alpha}$ 
represents a sum over the five direct amplitudes, while
$\sum^{5}_{\alpha=1} (-1)^{\alpha} 
F_{\alpha}(u,t){\stackrel{\textstyle \sim}{G}}_{\alpha}$
represents a sum over the five exchange amplitudes multiplied by
the sign factor arising for two nucleons.  The five direct
amplitudes are depicted in Fig.\ 1a and the five exchange amplitudes
are exhibited in Fig.\ 1b.  These ten elastic-scattering diagrams
will be used as source graphs to generate bremsstrahlung diagrams.

The Pauli principle imposes some restrictions on $F_{\alpha}(u,t)$.
For isotopic triplet states, we require that
\begin{equation}
   F_{\alpha}(u,t) = (-1)^{\alpha+1} F_{\alpha}(t,u) \ .
\end{equation}
If we interchange $\bar{q}^{\mu}_f$ with $\bar{p}^{\mu}_f$
(or $q^\mu_i$ with $p^\mu_i$), then ($i$) $u$ is interchanged
with $t$; ($ii$) $G_{\alpha}$ is interchanged with
${\stackrel{\textstyle \sim}{G}}_{\alpha}$; and ($iii$) the direct 
amplitude $F_{\alpha}(u,t) G_{\alpha}$ will be interchanged with the
exchange amplitude 
[$-(-1)^{\alpha} F_{\alpha}(u,t){\stackrel{\textstyle \sim}{G}}_{\alpha}$]
but with opposite sign.  Thus, the amplitude $F$ given by Eq.\ (2)
changes sign, and the Pauli principle is therefore satisfied.

\pagebreak
\section{Proton-proton bremsstrahlung amplitudes}
%---------------------------------------------------------
\subsection{External amplitudes}
%---------------------------------------------------------
We can use Figs.\ 1a and 1b as source graphs to generate
external emission $pp$ bremsstrahlung diagrams.

If the photon is emitted from the $q_f$--leg,
then we obtain Figs.\ 2a and 2b.  The amplitudes
corresponding to these two diagrams can be written as
\begin{eqnarray}
  M^{q_f}_{\mu}(u_1,t_p,\Delta_{q_f}) & = &
    e\sum^5_{\alpha=1} F_{\alpha}(u_1,t_p,\Delta_{q_f})
    \left[ \bar{u}(q_f)\Gamma_\mu
    \frac{1}{{\slash\hspace{-2mm}q}_f+{\slash\hspace{-2.5mm}K}-m}
    \lambda_\alpha u(q_i)\:\bar{u}(p_f)\lambda^\alpha u(p_i)
    \right. \nonumber \\ & & \left.
    +(-1)^\alpha \bar{u}(p_f)\lambda_\alpha u(q_i)\:\bar{u}(q_f)
    \Gamma_\mu
    \frac{1}{{\slash\hspace{-2mm}q}_f+{\slash\hspace{-2.5mm}K}-m}
    \lambda^\alpha u(p_i) \right] \ ,
\end{eqnarray}
where
\begin{eqnarray}
            u_1 & = & (p_f-q_i)^2 = (p_i-q_f-K)^2 \ , \nonumber \\
            t_p & = & (p_f-p_i)^2 = (q_i-q_f-K)^2 \ , \nonumber \\
   \Delta_{q_f} & = & (q_f+K)^2   = m^2 + 2q_f \cdot K \ , \nonumber
\end{eqnarray}
and
\begin{equation}
  \Gamma_\mu  = \gamma_\mu-i\frac{\kappa}{2m}\sigma_{\mu\nu}K^\nu
\end{equation}
is the electromagnetic vertex.  Here $e>0$ is the proton charge,
$\kappa$ is the anomalous magnetic moment of the proton, and we
have used three-body energy-momentum conservation for the $pp\gamma$
process,
\begin{eqnarray}
   q^\mu_i + p^\mu_i = q^\mu_f + p^\mu_f + K^\mu \ .
\end{eqnarray}
It is easy to show that
\begin{mathletters}
\begin{equation}
   \bar{u}(q_f)\Gamma_\mu \frac{1}{{\slash\hspace{-2mm}q}_f +
      {\slash\hspace{-2.5mm}K}-m} = \bar{u}(q_f)\left( \frac{q_{f\mu} +  
      R^{q_f}_\mu}{q_f \cdot K} \right) \ , \label{eq:11a}
\end{equation}
where 
\begin{equation}
   R^{q_f}_{\mu} = \frac{1}{4} \left[{\gamma_\mu}
      ,{\slash\hspace{-2.5mm}K}\right] + \frac{\kappa}{8m}
       \{ \left[{\gamma_\mu}
      ,{\slash\hspace{-2.5mm}K}\right],{\slash\hspace{-2mm}q}_f \}
       \ , \label{eq:11b} 
\end{equation}
\end{mathletters}

\noindent and we have used $[A,B] \equiv AB-BA$ and $\{A,B\} \equiv AB+BA$.
If we expand $F_{\alpha}(u_1,t_p,\Delta_{q_f})$ about $\Delta_{q_f}=m^2$,
\begin{equation}
   F_{\alpha}(u_1,t_p,\Delta_{q_f}) = \left. F_{\alpha}(u_1,t_p) +
   \frac{\partial F_{\alpha}(u_1,t_p,\Delta_{q_f})}{\partial\Delta_{q_f}} 
   \right|_{\Delta_{q_f}=m^2} (2q_f\cdot K) + \cdots \ ,
\end{equation}
where
\begin{eqnarray}
   F_{\alpha}(u_1,t_p) \equiv F_{\alpha}(u_1,t_p,m^2) \ , \nonumber
\end{eqnarray}
then Eq.\ (8) becomes
\begin{eqnarray}
  M^{q_f}_{\mu}(u_1,t_p,\Delta_{q_f}) & = &
    e\sum^5_{\alpha=1} \left[\left. F_{\alpha}(u_1,t_p) + (2q_f\cdot K)
    \frac{\partial F_{\alpha}(u_1,t_p,\Delta_{q_f})}{\partial\Delta_{q_f}}
    \right|_{\Delta_{q_f}=m^2} + \cdots \right] \nonumber \\ & & \times
    \left[\bar{u}(q_f) \left(\frac{q_{f\mu}+R^{q_f}_\mu}
    {q_f\cdot K}\right)
    \lambda_\alpha u(q_i)\:\bar{u}(p_f)\lambda^\alpha u(p_i) \right.
    \nonumber \\  & &
    + \left. (-1)^\alpha \bar{u}(p_f)\lambda_\alpha u(q_i)\:\bar{u}(q_f)
    \left(\frac{q_{f\mu}+R^{q_f}_\mu}{q_f\cdot K}\right)\lambda^\alpha
    u(p_i) \right] \ .
\end{eqnarray}

If the photon is emitted from the $q_i$--leg, then we get
Figs.\ 2c and 2d, and the corresponding amplitudes have the form
\begin{eqnarray}
  M^{q_i}_{\mu}(u_2,t_p,\Delta_{q_i}) & = &
    e\sum^5_{\alpha=1} F_{\alpha}(u_2,t_p,\Delta_{q_i})
    \left[ \bar{u}(q_f)\lambda_\alpha
    \frac{1}{{\slash\hspace{-2mm}q}_i-{\slash\hspace{-2.5mm}K}-m}
    \Gamma_\mu u(q_i)\:\bar{u}(p_f)\lambda^\alpha u(p_i)
    \right. \nonumber \\ & & \left.
    +(-1)^\alpha \bar{u}(p_f)\lambda_\alpha
    \frac{1}{{\slash\hspace{-2mm}q}_i-{\slash\hspace{-2.5mm}K}-m}
    \Gamma_\mu u(q_i)\:\bar{u}(q_f)\lambda^\alpha u(p_i) \right] \ ,
\end{eqnarray}
where
\begin{eqnarray}
   u_2 = (q_f-p_i)^2 = (q_i-p_f-K)^2 \ , \nonumber
\end{eqnarray}
and
\begin{eqnarray}
   \Delta_{q_i} = (q_i-K)^2 = m^2-2q_i\cdot K \ . \nonumber
\end{eqnarray}
If we use the relation
\begin{equation}
   \frac{1}{{\slash\hspace{-2mm}q}_i-{\slash\hspace{-2.5mm}K}-m}
   \Gamma_{\mu} u(q_i) = -\left(
   \frac{q_{i\mu}+R^{q_{i}}_\mu}{q_i\cdot K}\right) u(q_i) \ ,
\end{equation}
where $R^{q_i}_\mu$ is given by the same expression as
Eq.\ (11b) but with $q_f$ replaced by $q_i$, and expand
$F_\alpha(u_2,t_p,\Delta_{q_{i}})$ about $\Delta_{q_i}=m^2$,
\begin{equation}
   F_{\alpha}(u_2,t_p,\Delta_{q_i}) = \left. F_{\alpha}(u_2,t_p) +
   \frac{\partial F_{\alpha}(u_2,t_p,\Delta_{q_i})}{\partial\Delta_{q_i}} 
   \right|_{\Delta_{q_i}=m^2} (-2q_i\cdot K) + \cdots \ ,
\end{equation}
where
\begin{eqnarray}
   F_{\alpha}(u_2,t_p) \equiv F_{\alpha}(u_2,t_p,m^2) \ , \nonumber
\end{eqnarray}
we obtain from Eq.\ (14)
\begin{eqnarray}
  M^{q_i}_{\mu}(u_2,t_p,\Delta_{q_i}) & = &
   -e\sum^5_{\alpha=1} \left[\left. F_{\alpha}(u_2,t_p) - (2q_i\cdot K)
    \frac{\partial F_{\alpha}(u_2,t_p,\Delta_{q_i})}{\partial\Delta_{q_i}}
    \right|_{\Delta_{q_i}=m^2} + \cdots \right] \nonumber \\ & & \times
    \left[\bar{u}(q_f)\lambda_\alpha \left(\frac{q_{i\mu}+R^{q_i}_\mu}
    {q_i\cdot K}\right)
    u(q_i)\:\bar{u}(p_f)\lambda^\alpha u(p_i) \right.
    \nonumber \\  & &
    + \left. (-1)^\alpha \bar{u}(p_f)\lambda_\alpha
    \left(\frac{q_{i\mu}+R^{q_i}_\mu}{q_i\cdot K}\right)
    u(q_i)\:\bar{u}(q_f)\lambda^\alpha u(p_i) \right] \ .
\end{eqnarray}

Similarly, if the photon is emitted from the $p_f$--leg and $p_i$--leg,
then we obtain Figs.\ 2e and 2f and Figs.\ 2g and 2h, respectively.
The amplitudes corresponding to these figures have the following
expressions:
\begin{eqnarray}
  M^{p_f}_{\mu}(u_2,t_q,\Delta_{p_f}) & = &
    e\sum^5_{\alpha=1} \left[\left. F_{\alpha}(u_2,t_q) + (2p_f\cdot K)
    \frac{\partial F_{\alpha}(u_2,t_q,\Delta_{p_f})}{\partial\Delta_{p_f}}
    \right|_{\Delta_{p_f}=m^2} + \cdots \right] \nonumber \\ & & \times
    \left[\bar{u}(q_f)\lambda_\alpha u(q_i)\:\bar{u}(p_f)
    \left(\frac{p_{f\mu}+R^{p_f}_\mu}{p_f\cdot K}\right)
    \lambda^\alpha u(p_i) \right.
    \nonumber \\  & &
    + \left. (-1)^\alpha \bar{u}(p_f)
    \left(\frac{p_{f\mu}+R^{p_f}_\mu}{p_f\cdot K}\right)
    \lambda_\alpha u(q_i)\:\bar{u}(q_f)\lambda^\alpha u(p_i) \right] \ ,
\end{eqnarray}
and
\begin{eqnarray}
  M^{p_i}_{\mu}(u_1,t_q,\Delta_{p_i}) & = &
   -e\sum^5_{\alpha=1} \left[\left. F_{\alpha}(u_1,t_q) - (2p_i\cdot K)
    \frac{\partial F_{\alpha}(u_1,t_q,\Delta_{p_i})}{\partial\Delta_{p_i}}
    \right|_{\Delta_{p_i}=m^2} + \cdots \right] \nonumber \\ & & \times
    \left[\bar{u}(q_f)\lambda_\alpha u(q_i)\:\bar{u}(p_f)
    \lambda^\alpha \left(\frac{p_{i\mu}+R^{p_i}_\mu}{p_i\cdot K}\right)
    u(p_i) \right.
    \nonumber \\  & &
    + \left. (-1)^\alpha \bar{u}(p_f) \lambda_\alpha u(q_i)\:\bar{u}(q_f)
    \lambda^\alpha \left(\frac{p_{i\mu}+R^{p_i}_\mu}{p_i\cdot K}\right)
    u(p_i) \right] \ .
\end{eqnarray}
Here,
\begin{eqnarray}
            t_q & = & (q_f-q_i)^2 = (p_i-p_f-K)^2 \ , \nonumber \\
   \Delta_{p_f} & = & (p_f+K)^2 = m^2 + 2p_f \cdot K \ , \nonumber \\
   \Delta_{p_i} & = & (p_i-K)^2 = m^2 - 2p_i \cdot K \ ,
\end{eqnarray}
and $R^{p_f}_\mu$ and $R^{p_i}_\mu$ are given by the same expressions 
as $R^{q_f}_\mu$ in Eq.\ (11b) but with $q_f$ replaced by $p_f$ and
$p_i$, respectively.

The external emission process is the sum of emission processes from 
the four proton legs.  Therefore, the external bremsstrahlung
amplitude, $M^E_\mu$, can be written as
\begin{eqnarray}
   M^E_\mu & = &   M^{q_f}_\mu (u_1,t_p,\Delta_{q_f}) +
                   M^{q_i}_\mu (u_2,t_p,\Delta_{q_i}) \nonumber \\
           &   & + M^{p_f}_\mu (u_2,t_q,\Delta_{p_f}) +
                   M^{p_i}_\mu (u_1,t_q,\Delta_{p_i}) \ .
\end{eqnarray}

\subsection{Internal amplitudes}
%---------------------------------------------------------
The internal bremsstrahlung amplitude, $M^I_{\mu}$, can be
obtained from the gauge-invariance condition,
\begin{equation}
   (M^E_\mu + M^I_\mu) K^\mu = 0 \ .
\end{equation}
However, this condition alone cannot give a unique expression
for the amplitude $M^I_{\mu}$.  The ambiguity can be removed
if the additional requirement of satisfying the Pauli principle
is also imposed.
Because $R^Q_\mu$ $(Q=q_f,p_f,q_i,p_i)$ are separately gauge
invariant, {\it viz.} $R^Q_\mu K^\mu = 0$, we find
\begin{eqnarray}
   M^I_\mu K^{\mu} & = & - M^E_\mu K^\mu \nonumber \\ & = &
     -e\sum^{5}_{\alpha=1} \left[ F_{\alpha}(u_1,t_p)-
     F_{\alpha}(u_2,t_p)+F_{\alpha}(u_2,t_q)-F_{\alpha}(u_1,t_q) \right.
     \nonumber \\ & &
     \left. + (2q_f\cdot K)\frac{\partial
     F_{\alpha}(u_1,t_p,\Delta_{q_f})}{\partial\Delta_{q_f}}
     \right|_{\Delta_{q_f}=m^2}
     \left. + (2q_i\cdot K)\frac{\partial
     F_{\alpha}(u_2,t_p,\Delta_{q_i})}{\partial\Delta_{q_i}}
     \right|_{\Delta_{q_i}=m^2} \nonumber \\ & &
     \left. \left. + (2p_f\cdot K)\frac{\partial
     F_{\alpha}(u_2,t_q,\Delta_{p_f})}{\partial\Delta_{p_f}}
     \right|_{\Delta_{p_f}=m^2}
     + (2p_i\cdot K)\frac{\partial
     F_{\alpha}(u_1,t_q,\Delta_{p_i})}{\partial\Delta_{p_i}}
     \right|_{\Delta_{p_i}=m^2} \nonumber \\ & & \left. + \cdots \right]
     \left[ \bar{u}(q_f)\lambda_{\alpha}u(q_i)\:\bar{u}(p_f)
     \lambda^{\alpha}u(p_i) + (-1)^{\alpha} \bar{u}(p_f)
     \lambda_{\alpha}u(q_i)\:\bar{u}(q_f)\lambda^{\alpha}u(p_i)\right] \ .
\end{eqnarray}
Let us define
\begin{mathletters}
\begin{eqnarray}
   I_{\alpha} &\equiv&
      F_{\alpha}(u_1,t_p)-F_{\alpha}(u_2,t_p) +  
      F_{\alpha}(u_2,t_q)-F_{\alpha}(u_1,t_q) \\ \label{eq:24a} & = &
      \frac{1}{2}\left\{\left[F_{\alpha}(u_1,t_p)-F_{\alpha}(u_2,t_p)  
        \right]-\left[F_{\alpha}(u_1,t_q)-F_{\alpha}(u_2,t_q)\right] 
        \right.  \nonumber \\ & &
      + \left.\left[F_{\alpha}(u_1,t_p)-F_{\alpha}(u_1,t_q)\right] - 
        \left[F_{\alpha}(u_2,t_p)-F_{\alpha}(u_2,t_q)\right]\right\}
        \ . \label{eq:24b}
\end{eqnarray}
\end{mathletters}

\noindent The choice of the expression given by Eq.~(24b) is guided by
the requirement that the Pauli principle be satisfied.
Using the kinematic identities
\begin{eqnarray}
   u_1-u_2 & = & 2(q_f-p_i)\cdot K = 2(q_i-p_f)\cdot K \ , \nonumber \\
   t_p-t_q & = & 2(q_f-q_i)\cdot K = 2(p_i-p_f)\cdot K \ ,
\end{eqnarray}
and the mean-value theorem, we obtain
\begin{mathletters}
\begin{eqnarray}
   F_{\alpha}(u_1,t_p)-F_{\alpha}(u_2,t_p) & = & 2(q_i-p_f)\cdot K\;
      \frac{\partial F_{\alpha}(u_m,t_p)}{\partial u_m} \ ,
   \label{eq:26a} \\
   F_{\alpha}(u_1,t_q)-F_{\alpha}(u_2,t_q) & = & 2(q_i-p_f)\cdot K\;
      \frac{\partial F_{\alpha}(u^\prime_m,t_q)}
           {\partial u^\prime_m} \ ,
   \label{eq:26b} \\
   F_{\alpha}(u_1,t_p)-F_{\alpha}(u_1,t_q) & = & 2(p_i-p_f)\cdot K\;
      \frac{\partial F_{\alpha}(u_1,t_m)}{\partial t_m} \ ,
   \label{eq:26c} \\
   F_{\alpha}(u_2,t_p)-F_{\alpha}(u_2,t_q) & = & 2(p_i-p_f)\cdot K\;
        \frac{\partial F_{\alpha}(u_2,t^\prime_m)}
             {\partial t^\prime_m} \ ,
   \label{eq:26d}
\end{eqnarray}
\end{mathletters}

\noindent where $u_m$ and $u^\prime_m$ lie between $u_1$ and $u_2$,
and $t_m$ and $t^\prime_m$ lie between $t_p$ and $t_q$.
Inserting Eqs.\ (26a--26d) into Eq.\ (24b), we get
\begin{eqnarray}
   I_{\alpha} &=&
      (q_i-p_f)\cdot K\:\frac{\partial F_{\alpha}(u_m,t_p)}
      {\partial u_m} - (q_i-p_f)\cdot K\:\frac{\partial
      F_{\alpha}(u^\prime_m,t_q)}{\partial u^\prime_m}
      \nonumber \\ & &
      + (p_i-p_f)\cdot K\:\frac{\partial
      F_{\alpha}(u_1,t_m)}{\partial t_m} - (p_i-p_f)\cdot K\:
      \frac{\partial F_{\alpha}(u_2,t^\prime_m)}
      {\partial t^\prime_m} \ .
\end{eqnarray}
The expression for $M^I_\mu$ can now be generated if
we substitute Eq.\ (27) into Eq.\ (23).  We find
\begin{eqnarray}
   M^I_\mu & = & -e\sum^5_{\alpha=1} \left[
      (q_i-p_f)_\mu \frac{\partial F_{\alpha}(u_m,t_p)}
      {\partial u_m} - (q_i-p_f)_\mu \frac{\partial F_{\alpha}
      (u^\prime_m,t_q)}{\partial u^\prime_m}
      \right. \nonumber \\ & &
      + (p_i-p_f)_\mu \frac{\partial F_{\alpha}(u_1,t_m)}
      {\partial t_m} - (p_i-p_f)_\mu \frac{\partial F_{\alpha}
      (u_2,t^\prime_m)}{\partial t^\prime_m}
      \nonumber \\ & &
      + \left. 2q_{f\mu} \frac{\partial F_{\alpha}(u_1,t_p,\Delta_{q_f})}
      {\partial \Delta_{q_f}} \right|_{\Delta_{q_f}=m^2}
      + \left. 2q_{i\mu} \frac{\partial F_{\alpha}(u_2,t_p,\Delta_{q_i})}
      {\partial \Delta_{q_i}} \right|_{\Delta_{q_i}=m^2}
      \nonumber \\ & & \left.
      + \left. 2p_{f\mu} \frac{\partial F_{\alpha}(u_2,t_q,\Delta_{p_f})}
      {\partial \Delta_{p_f}} \right|_{\Delta_{p_f}=m^2}
      + \left. 2p_{i\mu} \frac{\partial F_{\alpha}(u_1,t_q,\Delta_{p_i})}
      {\partial \Delta_{p_i}} \right|_{\Delta_{p_i}=m^2}
      + \cdots \right] \nonumber \\ & &
      \left[ \bar{u}(q_f)\lambda_{\alpha}u(q_i)\:\bar{u}(p_f)
      \lambda^{\alpha}u(p_i) + (-1)^{\alpha}\bar{u}(p_f)\lambda_{\alpha}
      u(q_i)\:\bar{u}(q_f)\lambda^{\alpha}u(p_i) \right] \ .
\end{eqnarray}

\subsection{The Two-$u$-Two-$t$ special amplitude  
            $M^{TuTts}_\mu (u_1,u_2;t_p,t_q)$}
%---------------------------------------------------------
The amplitude $M^{TuTts}_{\mu}$ can be obtained if we combine
the amplitude $M^E_{\mu}$ given by Eq.\ (21) with the amplitude
$M^I_{\mu}$ given by Eq.\ (28),
\begin{eqnarray}
   M^T_\mu  & = &  M^E_\mu + M^I_\mu   \nonumber \\
            & = &  M^{TuTts}_\mu + {\cal O}(K) \ .
\end{eqnarray}
We observe that all off-shell derivative terms cancel precisely.
The derivatives of $F_{\alpha}$ with respect to
$u_m$, $u^\prime_m$, $t_m$, and
$t^\prime_m$ can be replaced by the finite differences
by using Eqs.\ (26a--26b).  For example, Eq.~(26a) gives
\begin{eqnarray}
  \frac{\partial F_{\alpha}(u_m,t_p)}{\partial u_m} & = &
  \frac{F_{\alpha}(u_1,t_p)-F_{\alpha}(u_2,t_p)}{2(q_i-p_f)\cdot K} \ .
  \nonumber
\end{eqnarray}
If we use the finite differences and the following relations
\begin{eqnarray}
   \frac{(q_f-p_i)\cdot\varepsilon}{(q_f-p_i)\cdot K} & = &
   \frac{(q_i-p_f)\cdot\varepsilon}{(q_i-p_f)\cdot K} \ , \nonumber \\
   \frac{(p_i-p_f)\cdot\varepsilon}{(p_i-p_f)\cdot K} & = &
   \frac{(q_f-q_i)\cdot\varepsilon}{(q_f-q_i)\cdot K} \ ,
\end{eqnarray}
the amplitude $M^{TuTts}_\mu$ can be written as
\begin{eqnarray}
   M^{TuTts}_\mu & = &  e\sum^5_{\alpha=1} \left[
   \bar{u}(q_f)X_{\alpha\mu}u(q_i)\:\bar{u}(p_f)\lambda^{\alpha}u(p_i)
  +\bar{u}(q_f)\lambda_{\alpha}u(q_i)\bar{u}(p_f)Y^{\alpha}_\mu u(p_i)
   \right. \nonumber \\ & & \left.
  +\bar{u}(p_f)\lambda^{\alpha}u(q_i)\bar{u}(q_f)Z_{\alpha\mu}u(p_i)
  +\bar{u}(p_f)T^{\alpha}_{\mu}u(q_i)\bar{u}(q_f)\lambda_{\alpha}u(p_i)
   \right] \ ,
\end{eqnarray}
where
\begin{eqnarray}
  X_{\alpha\mu}    & = & F_{\alpha}(u_1,t_p)\left[\frac{q_{f\mu}+
     R^{q_f}_{\mu}}{q_f\cdot K}-V_{\mu}\right]\lambda_{\alpha}
    -F_{\alpha}(u_2,t_p)\lambda_{\alpha}\left[\frac{q_{i\mu}+
     R^{q_i}_{\mu}}{q_i\cdot K}-V_{\mu}\right] \ , \nonumber \\
  Y^{\alpha}_{\mu} & = & F_{\alpha}(u_2,t_q)\left[\frac{p_{f \mu}+
     R^{p_f}_{\mu}}{p_f\cdot K}-V_{\mu}\right]\lambda^{\alpha}
    -F_{\alpha}(u_1,t_q)\lambda^{\alpha}\left[\frac{p_{i\mu}+
     R^{p_i}_{\mu}}{p_i\cdot K}-V_{\mu}\right] \ , \nonumber \\
  Z_{\alpha\mu}    & = &
               (-1)^{\alpha}F_{\alpha}(u_1,t_p)\left[\frac{q_{f\mu}+
     R^{q_f}_{\mu}}{q_f\cdot K}-V_{\mu}\right]\lambda_{\alpha}
-(-1)^{\alpha}F_{\alpha}(u_1,t_q)\lambda_{\alpha}\left[\frac{p_{i\mu}+
     R^{p_i}_{\mu}}{p_i\cdot K}-V_{\mu}\right] \ , \nonumber \\
  T^{\alpha}_{\mu} & = &
               (-1)^{\alpha}F_{\alpha}(u_2,t_q)\left[\frac{p_{f\mu}+
     R^{p_f}_{\mu}}{p_f\cdot K}-V_{\mu}\right]\lambda^{\alpha}
-(-1)^{\alpha}F_{\alpha}(u_2,t_p)\lambda^{\alpha}\left[\frac{q_{i\mu}+
     R^{q_i}_{\mu}}{q_i\cdot K}-V_{\mu}\right] \ ,
\end{eqnarray}
with
\begin{eqnarray}
   V_{\mu} & = &
   \frac{(q_f-p_i)_{\mu}}{2(q_f-p_i)\cdot K}+\frac{(q_f-q_i)_\mu}
   {2(q_f-q_i)\cdot K} = \frac{(q_i-p_f)_{\mu}}{2(q_i-p_f)
   \cdot K}+\frac{(p_i-p_f)_{\mu}}{2(p_i-p_f)\cdot K} \ . \nonumber
\end{eqnarray}
It is easy to verify that $M^{TuTts}_\mu$ is gauge invariant;
that is, one can demonstrate that $M^{TuTts}_\mu K^\mu = 0$.

If $p_i$ is interchanged with $q_i$, or if $q_f$ is interchanged 
with $p_f$, we find
\begin{eqnarray}
   X_{\alpha\mu}
     & \begin{array}[c]{c} \\ {\longleftrightarrow} \\
                           {q_i\leftrightarrow p_i} \end{array} &
  -Z_{\alpha\mu} \ , \nonumber \\
   Y^\alpha_\mu
     & \begin{array}[c]{c} \\ {\longleftrightarrow} \\
                           {q_i\leftrightarrow p_i} \end{array} &
  -T^\alpha_\mu \ , \nonumber \\
   X_{\alpha\mu}
     & \begin{array}[c]{c} \\ {\longleftrightarrow} \\
                           {q_f\leftrightarrow p_f} \end{array} &
  -T_{\alpha\mu} \ , \nonumber \\
   Y^\alpha_\mu
     & \begin{array}[c]{c} \\ {\longleftrightarrow} \\
                           {q_f\leftrightarrow p_f} \end{array} &
  -Z^\alpha_\mu \ .
\end{eqnarray}
Eq.\ (33) assures one that the amplitude $M^{TuTts}_{\mu}$
will change sign if $q_i \leftrightarrow p_i$ or
$q_f \leftrightarrow p_f$.  Hence, the Pauli principle is
still satisfied.

The amplitude $M^{TuTts}_{\mu}$ given by Eq.\ (31) can be separated
into an external contribution $M^{TuTts}_{\mu}(E)$ and an internal
contribution $M^{TuTts}_{\mu}(I)$,
\begin{equation}
   M^{TuTts}_{\mu} = M^{TuTts}_{\mu}(E) + M^{TuTts}_{\mu}(I) \ ,
\end{equation}
where
\begin{eqnarray}
   M^{TuTts}_{\mu}(E)  & = &  e\sum^5_{\alpha=1}\left[\bar{u}(q_f)  
     X_{\alpha\mu}(E)u(q_i)\:\bar{u}(p_f)\lambda^{\alpha}u(p_i)
    +\bar{u}(q_f)\lambda_{\alpha}u(q_i)\:\bar{u}(p_f)
     Y^{\alpha}_{\mu}(E)u(p_i)   \right. \nonumber \\ & & \left.
    +\bar{u}(p_f)\lambda^{\alpha}u(q_i)\:\bar{u}(q_f)Z_{\alpha\mu}(E)
     u(p_i)
    +\bar{u}(p_f)T^{\alpha}_{\mu}(E)u(q_i)\:\bar{u}(q_f)
    \lambda_{\alpha}u(p_i) \right] \ .
\end{eqnarray}
and
\begin{mathletters}
\begin{eqnarray}
   M^{TuTts}_{\mu}(I)  & = &
     -e\sum^{5}_{\alpha=1}V_{\mu}\left[F_{\alpha}(u_1,t_p)-
      F_{\alpha}(u_2,t_p)+F_{\alpha}(u_2,t_q)-F_{\alpha}(u_1,t_q)
   \right] \nonumber \\ & &
   \times \left[G_{\alpha}+(-1)^{\alpha}{\stackrel{\textstyle 
      \sim}{G}}_{\alpha}\right]   \label{eq:36a} \\   & = & 
    -e V_{\mu}\left[F(u_1,t_p)-F(u_2,t_p)+F(u_2,t_q)-F(u_1,t_q)\right]
            \ .   \label{eq:36b}
\end{eqnarray}
\end{mathletters}
In Eq.\ (35), $X_{\alpha\mu}(E)$, $Y^{\alpha}_{\mu}(E)$,
$Z_{\alpha\mu}(E)$,  and $T^{\alpha}_{\mu}(E)$
are given by the following expressions:
\begin{eqnarray}
  X_{\alpha\mu}(E)    & = & F_{\alpha}(u_1,t_p)\left(\frac{q_{f\mu}+
     R^{q_f}_{\mu}}{q_f\cdot K}\right)\lambda_{\alpha}
    -F_{\alpha}(u_2,t_p)\lambda_{\alpha}\left(\frac{q_{i\mu}+
     R^{q_i}_{\mu}}{q_i\cdot K}\right) \ , \nonumber \\
  Y^{\alpha}_{\mu}(E) & = & F_{\alpha}(u_2,t_q)\left(\frac{p_{f \mu}+
     R^{p_f}_{\mu}}{p_f\cdot K}\right)\lambda^{\alpha}
    -F_{\alpha}(u_1,t_q)\lambda^{\alpha}\left(\frac{p_{i\mu}+
     R^{p_i}_{\mu}}{p_i\cdot K}\right) \ , \nonumber \\
  Z_{\alpha\mu}(E)    & = &
               (-1)^{\alpha}F_{\alpha}(u_1,t_p)\left(\frac{q_{f\mu}+
     R^{q_f}_{\mu}}{q_f\cdot K}\right)\lambda_{\alpha}
-(-1)^{\alpha}F_{\alpha}(u_1,t_q)\lambda_{\alpha}\left(\frac{p_{i\mu}+
     R^{p_i}_{\mu}}{p_i\cdot K}\right) \ , \nonumber \\
  T^{\alpha}_{\mu}(E) & = &
               (-1)^{\alpha}F_{\alpha}(u_2,t_q)\left(\frac{p_{f\mu}+
     R^{p_f}_{\mu}}{p_f\cdot K}\right)\lambda^{\alpha}
-(-1)^{\alpha}F_{\alpha}(u_2,t_p)\lambda^{\alpha}\left(\frac{q_{i\mu}+
     R^{q_i}_{\mu}}{q_i\cdot K}\right) \ .
\end{eqnarray}
Note that we have used the definition of the elastic
amplitude, Eq.\ (2) (with $\bar{p}_f \to p_f$ and $\bar{q}_f \to q_f)$, 
to obtain Eq.\ (36b) from Eq.\ (36a).

The amplitude $M^{TuTts}_{\mu}(I)$ does not vanish in general.
If we use the following expansion,
\begin{eqnarray}
   F_{\alpha}(u_1,t_p) & - & F_{\alpha}(u_2,t_p)+F_{\alpha}(u_2,t_q)-
   F_{\alpha}(u_1,t_q) \nonumber \\ = & + &
   \left[2(q_i-p_f)\cdot K\right]\left[2(p_i-p_f)\cdot K\right]
   \frac{\partial^2 F_{\alpha}(u_1,t_q)}{\partial t_q\partial u_1}
   + {\cal O}(K^3) \ ,
\end{eqnarray}
then Eq.\ (36a) can be written as
\begin{eqnarray}
   M^{TuTts}_{\mu}(I) & = & -e\sum^5_{\alpha=1}\left[2(p_i-p_f)\cdot K
      \:(q_i-p_f)_{\mu} \right. \nonumber \\ & & \left.
      +2(q_i-p_f)\cdot K\:(p_i-p_f)_{\mu}\right]
      \frac{\partial^{2} F_{\alpha}(u_1,t_q)}{\partial t_q\partial u_1}
      \left[G_{\alpha}+(-1)^{\alpha}{\stackrel{\textstyle 
       \sim}{G}}_{\alpha}\right] + {\cal O}(K^2) \ ,
\end{eqnarray}
which shows that $M^{TuTts}_{\mu}(I)$ is order of $K$.  This
feature, together with the fact that $M^{TuTts}_{\mu}$ is free of
off-shell derivatives, proves that the amplitude $M^{TuTts}_{\mu}$ is
consistent with the soft-photon theorem.  In order to obey charge
conservation, the internal amplitude $M^{TuTts}_{\mu}(I)$ must
vanish at the tree level.  To see this, let us consider a 
one-boson-exchange (OBE) model.  For any OBE model, the elastic 
amplitudes $F(u_i,t_j)$ can be expressed as follows:
\begin{equation}
   F(u_i,t_j) = F_D(t_j) - F_E(u_i) \ , \;\;\; (i=1,2; \; j=p,q) \ .
\end{equation}
Here, $F_D(t_j)$ and $F_E(u_i)$ represent all direct amplitudes and 
all exchange amplitudes, respectively.
Inserting Eq.\ (40) into Eq.\ (36b), we find
\begin{equation}
   M^{TuTts}_{\mu}(I) = 0 \ .
\end{equation}

\pagebreak
\section{Discussion}
%---------------------------------------------------------
\subsection{The Two-$u$-Two-$t$ special amplitudes}
%---------------------------------------------------------
The amplitude $M^{TuTts}_{\mu}$ given by Eq.\ (31) is not the only
two-$u$-two-$t$ special amplitude which can be constructed from the
external amplitude Eq.\ (21) by imposing gauge invariance.  In
Eq.\ (23), the expression for $M^I_{\mu} K^{\mu}$ involves a factor
$I_\alpha$ defined by Eq.\ (24a).  If we rewrite $I_\alpha$ in the form
shown in Eq.\ (24b), and use the formulas given by Eqs.\ (26a--26d),
we obtain Eq.\ (27).  It is this expression for $I_\alpha$ which
gives us the amplitude $M^{TuTts}_{\mu}$.  This amplitude has many
good features: it is relativistic, gauge invariant, consistent with
the soft-photon theorem, and it satisfies the Pauli principle.

However, Eq.\ (27) is not a unique expression for $I_\alpha$.
If we substitute Eqs.\ (26a) and (26b) into Eq.\ (24a), we obtain
\begin{equation}
   \bar{I}_{\alpha} = 2(q_i-p_f)\cdot K\;\frac{\partial
      F_{\alpha}(u_m,t_p)}{\partial u_m}
 -2(q_i-p_f)\cdot K\;\frac{\partial F_{\alpha}(u^\prime_m,t_p)}
     {\partial u^\prime_m} \ , \nonumber
\end{equation}
which is different from Eq.\ (27).
If Eq.\ (42) were used, we would obtain a new amplitude,
$\bar{M}^{TuTts}_{\mu}$, which is given by the same expression
as $M^{TuTts}_{\mu}$ defined in Eqs.\ (31) and (32) but with
$V_{\mu}$ replaced by $\bar{V}_{\mu}$, with
\begin{equation}
   \bar{V}_{\mu} =
      \frac{(q_i-p_f)_{\mu}}{(q_i-p_f)\cdot K} =
      \frac{(q_f-p_i)_{\mu}}{(q_f-p_i)\cdot K} \ .
\end{equation}
If $p_i$ is interchanged with $q_i$ (or if $q_f$ is interchanged 
with $p_f$), then we have
\begin{mathletters}
\begin{eqnarray}
   \bar{V}_{\mu} \;\;
     \begin{array}[c]{c}                   \\
                         {\longrightarrow} \\
                         {q_i\leftrightarrow p_i} \end{array}
     \;\; \frac{(p_i-p_f)_{\mu}}{(p_i-p_f)\cdot K} =
          \frac{(q_f-q_i)_{\mu}}{(q_f-q_i)\cdot K}
      \neq \bar{V}_{\mu} \label{eq:44a}
\end{eqnarray}

\noindent while, on the other hand,
\begin{eqnarray}
   V_{\mu} \;\;
       \begin{array}[c]{c}                   \\
                           {\longrightarrow} \\
                           {q_i\leftrightarrow p_i} \end{array}
      \;\; V_{\mu} \ . \label{eq:44b}
\end{eqnarray}
\end{mathletters}

\noindent Thus, there is an important difference between the two amplitudes,
$M^{TuTts}_{\mu}$ and $\bar{M}^{TuTts}_{\mu}$, because $M^{TuTts}_{\mu}$
does satisfy the Pauli principle, while $\bar{M}^{TuTts}_{\mu}$
violates it.
This is the main reason why the amplitude $M^{TuTts}_{\mu}$, not
$\bar{M}^{TuTts}_{\mu}$, should be used for the $pp\gamma$ process.

If we apply the Fierz transformation,
\begin{equation}
   \left( \begin{array}{c}
    {\stackrel{\textstyle \sim}{G}}_1 \\ 
    {\stackrel{\textstyle \sim}{G}}_2 \\ 
    {\stackrel{\textstyle \sim}{G}}_3 \\ 
    {\stackrel{\textstyle \sim}{G}}_4 \\ 
    {\stackrel{\textstyle \sim}{G}}_5
     \end{array} \right) =
   \frac{1}{4} \left( \begin{array}{rrrrr}
               1  &  1  &  1  &  1  &  1  \\
               6  & -2  &  0  &  0  &  6  \\
               4  &  0  & -2  &  2  & -4  \\
               4  &  0  &  2  & -2  & -4  \\
               1  &  1  & -1  & -1  &  1  \end{array} \right)
   \left( \begin{array}{c}
      G_1 \\ G_2 \\ G_3 \\ G_4 \\ G_5 \end{array} \right) \ ,
\end{equation}
we can write Eq.\ (2) in the form
\begin{equation}
   F = \sum^5_{\alpha=1} F^e_{\alpha}(u,t) G_{\alpha} \ ,
\end{equation}
or in the form
\begin{equation}
   F = \sum^5_{\alpha=1} F^e_{\alpha}(s,t) G_{\alpha} \ ,
\end{equation}
where
\begin{equation}
   \left( \begin{array}{c}
    F^e_1 \\ F^e_2 \\ F^e_3 \\ F^e_4 \\ F^e_5
     \end{array} \right) =
   \frac{1}{4} \left( \begin{array}{rrrrr}
               3  &  6  & -4  &  4  & -1  \\
              -1  &  2  &  0  &  0  & -1  \\
              -1  &  0  &  6  &  2  &  1  \\
              -1  &  0  & -2  &  2  &  1  \\
              -1  &  6  &  4  & -4  &  3  \end{array} \right)
   \left( \begin{array}{c}
      F_1 \\ F_2 \\ F_3 \\ F_4 \\ F_5 \end{array} \right) \ .
\end{equation}
Equation (46) is obtained when we choose ($u,t$) to be the two independent
variables; i.e., we use $F_{\alpha}=F_{\alpha}(u,t)$ in Eq.\ (2).
On the other hand, Eq.\ (47)
is obtained if the two independent variables are ($s,t$).
For the $pp$ elastic case, the expressions for $F$ given by Eqs.\
(2), (46), and (47) are identical.  However, if these expressions are
used as input to generate $pp\gamma$ amplitudes, then the constructed
amplitudes will be different.  The amplitude generated from Eq.\ (47) will
be discussed in next subsection.  Here, we would like to present, without
showing the details of derivation, two more two-$u$-two-$t$ special
amplitudes which can be obtained from Eq.\ (46).  We have
\begin{equation}
   M^{TuTts}_{i\mu} = e\sum^5_{\alpha=1} \left[\bar{u}(q_f)
     X_{i\alpha\mu}u(q_i)\:\bar{u}(p_f)\lambda^{\alpha}u(p_i)
    +\bar{u}(q_f)\lambda_{\alpha}u(q_i)\:\bar{u}(p_f)
     Y^{\alpha}_{i\mu}u(p_i) \right] \ ,
\end{equation}
where ($i=1,2$),
\begin{eqnarray}
   X_{i\alpha\mu}    & = & F^e_{\alpha}(u_1,t_p) \left[\frac{q_{f\mu}+
   R^{q_f}_{\mu}}{q_f\cdot K}-V_{i\mu}\right]\lambda_{\alpha}
   -F^e_{\alpha}(u_2,t_p)\lambda_{\alpha}
   \left[\frac{q_{i\mu}+R^{q_i}_{\mu}}{q_i\cdot K}-V_{i\mu}\right]
   \ , \nonumber \\
   Y^{\alpha}_{i\mu} & = & F^e_{\alpha}(u_2,t_q) \left[\frac{p_{f\mu}+
   R^{p_f}_{\mu}}{p_f\cdot K}-V_{i\mu}\right]\lambda^{\alpha}
   -F^e_{\alpha}(u_1,t_q)\lambda^{\alpha} \left[\frac{p_{i\mu}+
   R^{p_{i}}_{\mu}}{p_i\cdot K}-V_{i\mu}\right] \ ,
\end{eqnarray}
with
\begin{eqnarray}
   V_{1\mu} & = & V_{\mu} \ , \nonumber \\
   V_{2\mu} & = & \bar{V}_{\mu} \ , \nonumber
\end{eqnarray}
and $V_{\mu}$ and $\bar{V}_{\mu}$ are defined by Eqs.\ (32) and (43),
respectively.  In deriving $M^{TuTts}_{1\mu}$, we have used Eqs.\
(24b) and (26a--26d).  On the other hand, we have used
Eqs.\ (24a), (26a), and (26b) to derive $M^{TuTts}_{2\mu}$.
It can be shown that the amplitude $M^{TuTts}_{1\mu}$ is identical
to the amplitude $M^{TuTts}_{\mu}$.  Let us outline the proof as
follows:  If we write Eqs.\ (45) and (48) in the form 
${\stackrel{\textstyle \sim}{G}}_{\alpha}$ $=$
$\sum_{\beta} C_{\alpha\beta} G_{\beta}$ and 
$F^e_{\alpha} = \sum_{\beta} \bar{C}_{\alpha\beta} F_{\beta}$,
respectively, then $C_{\alpha\beta}$ and $\bar{C}_{\alpha\beta}$
can be defined.  The first step is to transform the exchange terms,
the third and fourth terms of Eq.\ (31), into the same form as the
direct terms by using the Fierz identity, 
$(\lambda_\alpha)_{ab}(\lambda^\alpha)_{cd}$ $=$ 
$\sum^5_{\beta=1}C_{\alpha\beta}(\lambda_\beta)_{ad}(\lambda^\beta)_{cb}$.
The second step is to combine these transformed direct terms
obtained in the first step with the original direct terms of 
Eq.\ (31).  The amplitude $M^{TuTts}_{1\mu}$ can be easily obtained
if we use the identity $\bar{C}_{\alpha\beta}$ $=$ 
$\delta_{\alpha\beta}+ (-1)^{\beta} C_{\beta\alpha}$.  Clearly, 
$M^{TuTts}_{1\mu}$ satisfies the Pauli principle, because it is 
identical to $M^{TuTts}_{\mu}$.  The proof can also be carried out 
starting directly from $M^{TuTts}_{1\mu}$.  This can be accomplished 
by using another identity, 
$\sum^5_{\alpha=1} C_{\alpha\gamma} \bar{C}_{\alpha\beta}$
$=$  $(-1)^\beta \bar{C}_{\gamma\beta}$.

The amplitude $M^{TuTts}_{2\mu}$ ($\equiv \bar{M}^{TuTts}_{\mu}$) 
has been used in Ref.~\cite{Lio95}.
This amplitude violates the Pauli principle because its internal
amplitude depends upon $\bar{V}_\mu$.  However, the violation is only 
of order $K$.  To see this, one need only observe that the internal 
amplitude for $M^{TuTts}_{2\mu}$ is given by the same expresssion
as that in Eq.\ (36a) but with $V_\mu$ replaced by $\bar{V}_\mu$,
$F_\alpha$ replaced by $F^e_\alpha$, and the 
${\stackrel{\textstyle \sim}{G}}_{\alpha}$ omitted.  If one then
carries out an expansion similar to that given by Eq.\ (38), one
sees that the internal amplitude contributes only to the term of
order $K$, the third term, in the soft-photon expansion.

The amplitudes $M^{TuTts}_{\mu}$ and $M^{TuTts}_{2\mu}$
have been numerically studied.  We found that the 
$pp\gamma$ cross sections calculated from the two amplitudes are 
not significantly different, except for those cases when both proton
scattering angles are very small and the photon angle $\psi_\gamma$
is around 180$^o$.  The amplitude $M^{TuTts}_{\mu}$ gives the
expected symmetric angular distribution for 
$0 \leq \psi_\gamma \leq 180^o$ and $180^o \leq \psi_\gamma \leq 360^o$,
while $M^{TuTts}_{2\mu}$ yields angular distributions which are 
slightly distorted around the point $\psi_\gamma = 180^o$.  
Otherwise, both calculated corss sections are in good agreement
with the experimental data and most of the potential model
predictions.

Finally, it should be pointed out that the Low amplitude can
be derived from either $M^{TuTts}_{\mu}$ or $M^{TuTts}_{1\mu}$,
and therefore it satisfies the Pauli principle.

\subsection{The Two-$s$-Two-$t$ special amplitudes  
            $M^{TsTts}_{\mu}(s_i,s_f;t_p,t_q)$}
%---------------------------------------------------------
Another class of amplitudes, the two-$s$-two-$t$ special amplitudes,
can be derived if ($s,t$) are chosen to be the independent variables.
The input (elastic-scattering amplitude) used to generate this class of
amplitudes can be either Eq.\ (2) with $F_{\alpha}=F_{\alpha}(s,t)$
(without introducing the Fierz transformation) or Eq.\ (47), which is
obtained from Eq.\ (2) by applying the Fierz transformation. 
In other words, two amplitudes can be constructed, but it can be
shown that they are identical.  The same procedure as outlined in
the previous subsection can be followed to obtain the proof. 
Here, we will just
present the expressions for these two amplitudes without derivation,
because the procedures for deriving them are very similar to those
used in the previous sections for $M_\mu^{TuTts}$.

If Eq.\ (2) with $F_{\alpha}=F_{\alpha}(s,t)$ is used as input,
the resulting $pp\gamma$ amplitude assumes the form
\begin{eqnarray}
   M^{TsTts}_{1\mu}(s_i,s_f;t_p,t_q) & = &
     e\sum^5_{\alpha=1} \left[\bar{u}(q_f)
     {\stackrel{\textstyle \sim}{X}}_{1\alpha\mu}u(q_i)\:
     \bar{u}(p_f)\lambda^{\alpha}u(p_i)
     +\bar{u}(q_f)\lambda_{\alpha}u(q_i)\:\bar{u}(p_f)
     {\stackrel{\textstyle \sim}{Y}}^{\,\alpha}_{1\mu}u(p_i) 
      \right. \nonumber \\ & & \left.
     +\bar{u}(p_f)\lambda_{\alpha}u(q_i)\:\bar{u}(q_f)
     {\stackrel{\textstyle \sim}{Z}}^{\,\alpha}_{1\mu}u(p_i)\:
     +\bar{u}(p_f){\stackrel{\textstyle 
      \sim}{T}}_{1\alpha\mu}u(q_i)\:\bar{u}(q_f)
     \lambda^{\alpha}u(p_i)\right] \ , 
\end{eqnarray} 
where
\begin{eqnarray}
   {\stackrel{\textstyle \sim}{X}}_{1\alpha\mu}  & = &
   F_{\alpha}(s_i,t_p)\left(\frac{q_{f\mu}+
   R^{q_{f}}_{\mu}}{q_{f}\cdot K}-W_{\mu}\right)\lambda_{\alpha}
   -F_{\alpha}(s_f,t_p)\lambda_{\alpha}\left(\frac{q_{i\mu}+
   R^{q_{i}}_{\mu}}{q_{i}\cdot K}-W_{\mu}\right) \ , \nonumber \\
   {\stackrel{\textstyle \sim}{Y}}^{\,\alpha}_{1 \mu} & = &
   F_{\alpha}(s_i,t_q)\left(\frac{p_{f\mu}+
   R^{p_{f}}_{\mu}}{p_{f}\cdot K}-W_{\mu}\right)\lambda^{\alpha}
   -F_{\alpha}(s_f,t_q)\lambda^{\alpha}\left(\frac{p_{i\mu}+
   R^{p_{i}}_{\mu}}{p_{i}\cdot K}-W_{\mu}\right) \ , \nonumber \\
   {\stackrel{\textstyle \sim}{Z}}^{\,\alpha}_{1\mu} & = &
   (-1)^{\alpha}F_{\alpha}(s_i,t_p)
   \left(\frac{q_{f\mu}+R^{q_{f}}_{\mu}}{q_{f}\cdot K}-W_{\mu}\right)
   \lambda^{\alpha}
   -(-1)^{\alpha}F_{\alpha}(s_f,t_q)\lambda^{\alpha}\left(\frac{p_{i\mu}
   +R^{p_{i}}_{\mu}}{p_{i}\cdot k}-W_{\mu}\right) \ , \nonumber \\
   {\stackrel{\textstyle \sim}{T}}_{1\alpha\mu} & = &
   (-1)^{\alpha}F_{\alpha}(s_i,t_q) \left(\frac{p_{f\mu}
   +R^{p_{f}}_{\mu}}{p_{f}\cdot K}-W_{\mu}\right)\lambda_{\alpha}
   -(-1)^{\alpha}F_{\alpha}(s_f,t_p)\lambda_{\alpha}\left(\frac{q_{i\mu}
   +R^{q_{i}}_{\mu}}{q_{i}\cdot K}-W_{\mu}\right) \ ,
\end{eqnarray}
with
\begin{eqnarray}
   W_{\mu} & = & \frac{(p_i+q_i)_\mu}{(p_i+q_i)\cdot K} =
             \frac{(p_f+q_f)_\mu}{(p_f+q_f)\cdot K} \ . \nonumber
\end{eqnarray}
On the other hand, if Eq.\ (47) is used to generate the
two-$s$-two-$t$ special amplitude, we obtain
\begin{equation}
   M^{TsTts}_{2\mu}(s_i,s_f;t_p,t_q) =
      e\sum^5_{\alpha=1} \left[\bar{u}(q_f)
      {\stackrel{\textstyle \sim}{X}}_{2\alpha\mu}u(q_i)
      \:\bar{u}(p_f)\lambda^{\alpha}u(p_i)
      +\bar{u}(q_f)\lambda_{\alpha}u(q_i)\:\bar{u}(p_f)
      {\stackrel{\textstyle \sim}{Y}}^{\,\alpha}_{2\mu}u(p_i)\right] \ ,
\end{equation}
where
\begin{eqnarray}
   {\stackrel{\textstyle \sim}{X}}_{2\alpha\mu} & = &
      F^e_{\alpha}(s_i,t_p)\left(\frac{q_{f\mu}+
      R^{q_{f}}_{\mu}}{q_f\cdot K}-W_{\mu}\right)\lambda_{\alpha}
      -F^e_{\alpha}(s_f,t_p)\lambda_{\alpha}\left(\frac{q_{i\mu}+
      R^{q_{i}}_{\mu}}{q_i\cdot K}-W_{\mu}\right) \ , \nonumber \\
   {\stackrel{\textstyle \sim}{Y}}^{\,\alpha}_{2\mu}  & = &
      F^e_{\alpha}(s_i,t_q)\left(\frac{p_{f\mu}+
      R^{p_{f}}_{\mu}}{p_f\cdot K}-W_{\mu}\right)\lambda^{\alpha}
      -F^e_{\alpha}(s_f,t_q)\lambda^{\alpha}\left(\frac{p_{i\mu}+
      R^{p_i}_{\mu}}{p_i\cdot K}-W_{\mu}\right) \ .
\end{eqnarray}
Obviously, both amplitudes ($M^{TsTts}_{1\mu} \equiv M^{TsTts}_{2\mu}$)
are relativistic, gauge invariant, and consistent with the soft-photon
theorem.  The most serious theoretical arguments against the use of these
two amplitudes to describe the $pp\gamma$ process are that they are quite
different from the amplitude constructed from the OBE model and that they
violate the Pauli principle.  If $q_i$ is interchanged with $p_i$ (or 
$q_f$ with $P_f$), then $t_p$ and $t_q$ will be transformed into $u_1$
and $u_2$, and one obtains the amplitude $M^{TsTts}_{i\mu}(s_i,s_f;u_1,u_2)$ 
$(i=1,2)$, which is completely different from the amplitude 
$-M^{TsTts}_{i\mu}(s_i,s_f;t_p,t_q)$.

We have shown that $M^{TuTts}_{\mu}$ (or $M^{TuTts}_{1\mu}$) is a 
suitable amplitude to use in describing the $pp\gamma$ process, because 
it meets all theoretical requirements.  As we have noted above, even 
though the amplitude $M^{TuTts}_{2\mu}$ does not satisfy the Pauli 
principle at the order $K$, its
numerical predictions are close to the results calculated from
the amplitude $M^{TuTts}_{\mu}$, potential 
models~\cite{Wor86,Bro91,Her92,Jet93,Kat93}, and the OBE 
model~\cite{Lio95a}, for most cases.  That is, the violation
of the Pauli principle is not serious, and the amplitude describes
the $pp\gamma$ cross sections rather well.  The ($s,t$) class of
amplitudes ($M^{TsTts}_{1\mu} \equiv M^{TsTts}_{2\mu}$), on the
other hand, {\it cannot} reproduce the OBE result.  The OBE
amplitude, in fact, belongs to the ($u,t$) class of amplitudes.
Moreover, the violation of the Pauli principle for the ($s,t$)
class of amplitudes is far more serious than it is for the ($u,t$)
class of amplitudes.
This is the most compelling reason why the ($u,t$) class of
amplitudes should be used to describe the $pp\gamma$ process, and
why the optimal amplitude is $M^{TuTts}_{\mu}$ given by Eq.\ (31).

\acknowledgments
The work of M.\ K.\ L.\ was supported
in part by the City University of New York Professional
Staff Congress-Board of Higher Education Faculty Research
Award Program, while the work of R.\ T.\ was included
in the research program of the Stichting voor Fundamenteel
Onderzoek der Materie (FOM) with financial support from the
Nederlandse Organisatie voor Wetenschappelijk Onderzoek (NWO),
and the work of B.\ F.\ G.\ was performed under the auspices
of the U.\ S.\ Department of Energy.  We thank A.\ Korchin and
O.\ Scholten for raising the question of how the Pauli principle
should be incorporated in the soft-photon approximation for
the case of identical fermions.  We thank H. Fearing for his
comments regarding the Low amplitude.  Helpful discussions with 
Yi Li are gratefully acknowledged.

\newpage
\vspace*{12pt}
\noindent Figure Captions

\vspace{12pt}
\noindent Fig.\ 1.  Schematic representation of the proton-proton elastic
scattering process: (a) corresponds to a sum over the five direct
amplitudes; (b) corresponds to a sum over the five exchange amplitudes
multiplied by the sign factor $(-1)^\alpha$.

\vspace{12pt}
\noindent Fig.\ 2. The external bremsstrahlung diagrams generated from
Fig.\ 1:  (a) and (b) represent photon emission from the $q_f$--leg;
(c) and (d) from the $q_i$--leg; (e) and (f) from the $p_f$--leg; (g)
and (h) from the $p_i$--leg.

\end{document}